\documentclass[twocolumn]{aastex6}
\begin{document}


\title{The Viewing Geometry of Brown Dwarfs Influences Their Observed Colours and Variability Amplitudes}



\author{Johanna M. Vos}
\affil{Institute for Astronomy, University of Edinburgh, Blackford Hill View, Edinburgh EH9 3HJ, UK; jvos@roe.ac.uk}

\author{Katelyn N. Allers}
\affil{Department of Physics and Astronomy, Bucknell University, Lewisburg, PA 17837, USA}

\author{Beth A. Biller}
\affil{Institute for Astronomy, University of Edinburgh, Blackford Hill View, Edinburgh EH9 3HJ, UK}



\begin{abstract}

In this paper we study the full sample of known \textit{Spitzer} [$3.6~\mu$m] and $J$-band variable brown dwarfs.{ We calculate the rotational velocities, $v\sin i$, of 16 variable brown dwarfs using archival Keck NIRSPEC data and compute the inclination angles of 19 variable brown dwarfs.}
 The results obtained show that all objects in the sample with mid-IR variability detections are inclined at an angle $>20^{\circ}$, while all objects in the sample displaying $J$-band variability have an inclination angle $>35^{\circ}$.
$J$-band variability appears to be more affected by inclination than \textit{Spitzer} [$3.6~\mu$m] variability, and is strongly attenuated at lower inclinations. Since $J$-band observations probe deeper into the atmosphere than mid-IR observations, this effect may be due to the increased atmospheric path length of $J$-band flux at lower inclinations.
 We find a statistically significant correlation between the colour anomaly and inclination of our sample, where field objects viewed equator-on appear redder than objects viewed at lower inclinations. 
Considering the full sample of known variable L, T and Y spectral type objects in the literature, we find that the variability properties of the two bands display notably different trends, due to both intrinsic differences between bands and the sensitivity of ground-based versus space-based searches. However, in both bands we find that variability amplitude may reach a maximum at $\sim7-9~$hr periods.
Finally, we find a strong correlation between colour anomaly and variability amplitude for both the $J$-band and mid-IR variability detections, where redder objects display higher variability amplitudes.

\end{abstract}




\section{Introduction} \label{sec:intro}


Time-resolved photometric variability monitoring is a
key probe of atmospheric structures in brown dwarf atmospheres, revealing a
periodic modulation of the lightcurve as a feature rotates in and out of
view. The combination of surface inhomogeneities in brown dwarf atmospheres and rapid rotation has long motivated searches for photometric variability in these objects. The first unambiguous detections \citep{Artigau2009, Radigan2012} were high-amplitude variable objects at the L/T transition. More recently, space and
ground-based surveys in the near-IR and mid- IR have revealed that variability is common across the
full range of L and T spectral types \citep{Wilson2014, Radigan2014, Buenzli2014, metchev2015a}. In fact, \citet{metchev2015a} concluded from a \textit{Spitzer} survey that most L and T spectral type brown dwarfs display low-level variability. To date, variability has been detected in $\sim30$ brown dwarfs, with $\sim5$ objects displaying high amplitude variability ($>5\%$).   Of the highest variability brown dwarfs discovered thus far, it
is known that WISE 1049B is viewed roughly equator-on \added{, with a viewing angle $i \geq 60^{\circ}$}
\citep{Crossfield2014}. For an equator-on object  (with an inclination
angle, $i \sim90^{\circ}$) we measure the full variability amplitude via
photometric monitoring. In contrast, we measure lower variability amplitudes for low-inclination objects \citep{Kostov2013}. In this paper, we aim to ascertain whether the range of observed amplitudes is due to properties intrinsic to each brown dwarf  or whether it can be explained by consideration of their inclination angles.

A proper motion survey conducted by \citet{Kirkpatrick2010} led to the discovery of a number of L spectral brown dwarfs that were redder than the median and L type brown dwarfs that were bluer than the median. Their kinematics revealed that they are both drawn from a relatively old population. This led to the possibility that both of these phenomena occur in the same objects, and that viewing angle determines their spectral appearance. This idea that spectral appearance is influenced by inclination angle is again suggested by \citet{metchev2015a}, who find a tentative correlation between near-IR colour and high-amplitude variability. If inclination angle affects the observed amplitude as well as the observed near-IR colour, then these two measurements will be related. The calculation of the inclination angle of brown dwarfs is critical in testing the relation between inclination and atmospheric appearance.


Attempts to model the cloud structure observed on variable brown dwarfs as patchy spots of thick and thin clouds have also been hindered by the unknown inclination of such objects.  \citet{Walkowicz2013} performed extended numerical experiments to assess degeneracies in models of spotted lightcurves, and confirmed that in the absence of inclination constraints, spot latitudes cannot be determined, regardless of data quality. \citet{Apai2013} obtained high-precision HST near-infrared spectroscopy of two highly variable L/T transition dwarfs 2M2139+02  and SIMP 0136. Surface brightness distributions were modelled using the inclination angle as an optimizable parameter, although the results are highly degenerate with respect to inclination, as multiple spot models with different inclinations fit the same lightcurve equally well. More recently, \citet{Karalidi2016} updated their \textit{Aeolus} routine, a Markov-Chain Monte Carlo code that can map the top-of-the-atmosphere structure of an ultracool atmosphere, to fit for inclination as a free parameter and successfully retrieved an inclination of $69\pm8^{\circ}$ for WISE 1049B, in agreement with the earlier measurement by \citet{Crossfield2014}. Constraining the inclination angles of variable brown dwarfs will allow us to model brown dwarf atmospheres in unprecedented detail.

In this paper we study the effects of inclination angle on the observed properties of brown dwarfs for the first time. We measure the rotational velocity, $v\sin i$, of 16 variable brown dwarfs (11 of which have no previous measurement in the literature) using archival Keck data, and use estimates of radius to determine their inclination angles. We investigate the relationship between inclination angle, variability amplitude and colour anomaly. Furthermore, we investigate the entire list of known brown dwarf $J$-band, \textit{Spitzer} [$3.6~\mu$m] and Kepler variability detections and explore the relations between variability amplitude, rotation period and colour anomaly. In Section \ref{sec:thesample} we discuss the sample of variable brown dwarfs. In Sections \ref{sec:data} --- \ref{sec:methods} we discuss the archival data and our methods in calculating inclinations. We discuss our results in Section \ref{sec:results}.


\section{The Sample}
\label{sec:thesample}

\begin{table*}[tb]
\centering
\caption{Variable brown dwarfs with known periods and archival spectra. Starred objects are those for which we adopted $v\sin i$  values from the literature. $(J-K_S)$ colours are $2MASS$. }
\label{tab:sample}
\begin{tabular*}{\textwidth}{lllllllll}

\hline \hline

\multicolumn{1}{l}{} & 
	 \multicolumn{1}{c}{}& 
		 \multicolumn{1}{c}{{[}3.6{]} Amp}& 
			 \multicolumn{1}{c}{$J$ Amp}  &
			 	\multicolumn{1}{c}{Kep Amp}&
					\multicolumn{1}{c}{Period }&
			 			\multicolumn{1}{c}{$v \sin i$}&
			 				\multicolumn{1}{c}{}&
								 \multicolumn{1}{c}{}\\
\multicolumn{1}{c}{ Name} & 
	 \multicolumn{1}{c}{ Spt}& 
		 \multicolumn{1}{c}{(\%)}& 
			 \multicolumn{1}{c}{(\%)}  &
			 	\multicolumn{1}{c}{(\%)}&
					\multicolumn{1}{c}{(hr)}&
			 			\multicolumn{1}{c}{(kms$^{-1})$}&
			 				\multicolumn{1}{c}{$(J-K_S)$ }&
								 \multicolumn{1}{c}{Refs}\\

\hline
2M0036+18  		&L3.5 	& $0.47\pm 0.05$      	& $1.22 \pm0.04$     		&         	& $2.7 \pm 0.3$    	&$35.12\pm0.57$				&1.41		& 1, 2, 3	\\
W0047			&L6		&					&$10$				&		&$13.2 \pm 0.14$	&$4.3\pm2.2$,$6.7^{+0.7}_{-1.4}$	&2.55		& 4, 5\\ 
2M0103+19		&L4		&$0.47\pm0.05$		&				&			&$2.7\pm0.3$&				&2.14					& 1 \\
2M0107+00   		&L8		& $1.27 \pm 0.13$     	&             			&         			& 5                			&				&2.11		& 1	\\
SIMP 0136 		&T2.5 	&$1.5\pm0.2$                   	& 5           			&         			& $2.414 \pm 0.078$ 	&				&0.90		& 6, 7	\\ 
SDSS0423-04		&T0		&					&$0.8\pm0.08$		&				&$2\pm0.4$			&				&1.54		&8	\\
WISE1049B*		&T0.5 	&					&7				&				&$4.87\pm0.01$		&$26.1\pm0.2$		&1.89		& 9, 10, 11	\\
DENIS 1058*		&L3		&$0.39\pm0.04$		&0.843			&				&$4.3\pm	0.31$		&$37.5\pm2.5$		&1.62		& 1, 12, 13	\\	
2M1126-50   		&L4.5  	&$0.21 \pm0.04$		& $1.2\pm0.1$        	&				& $3.2\pm0.3$   		&				&1.17		& 1, 6	\\
2M1507-16   		&L5		& $0.57   \pm0.04$     	&             			&         			& $2.5 \pm 0.1 $    		&$21.27\pm1.86$	&1.51		& 1, 3	 \\
2M1615+49   		&L4		& $0.9 \pm 0.2 $      		&             			&         			& 24                			&				&2.47		& 1	 \\
SIMP 1629		&T2		&                     			& 4.3       			&        			& $>7$              		&				&1.25		& 6	\\
2M1721+33   		&L3		& $0.33 \pm 0.07$    		&             			&         			& $2.6 \pm 0.1$    		&				&1.14		& 1	\\
2M1821+14   		&L4.5	& $0.54 \pm 0.05$    		&             			&         			& $4.2 \pm 0.1$     		&$28.85\pm 0.16$	&1.78		& 1, 3	\\
2M1906+40   		&L1  	 	&      					&             			& 1.5				& 8.9              			&$11.2\pm 2.2$		&1.31		& 14	\\
PSO-318*			&L7.5	&					&$10 \pm 1$		&				&$7.5 \pm 2.5$			&$17.5^{+2.3}_{-2.8}	$&2.78		&15, 16\\
2M2139+02   		&T1.5 	&   $11\pm1$           		& 26          		&         			& $7.618 \pm 0.178$ 	&				&1.68		& 6, 7	\\
2M2148+40		&L6		& $1.33 \pm 0.07$		&				&				&$19\pm4$			&				&2.38		& 1	\\
2M2208+29   		&L3		& $0.69\pm0.07$               &             			&         			& $3.5 \pm 0.2$    		&				&1.65		& 1 \\ 
	[1ex]
\hline
\end{tabular*}
\\[1.5ex]
{\bf References:} --- (1) \citet{metchev2015a}, (2) \citet{Croll2016}, (3) \citet{Blake2010}, (4) \citet{Lew2016}, (5) \citet{Gizis2015}, (6) \citet{Radigan2014}, (7) \citet{Yang2016}, (8) \citet{Clarke2008}, (9) \citet{Gillon2013},  (10) \citet{Biller2013}, (11) \citet{Crossfield2014},   (12) \citet{Heinze2014}, (13) \citet{Basri2000},  (14) \citet{Gizis2013},   (15) \citet{Biller2015}, (16) \citet{Allers2016}.

\end{table*}

\begin{table}[htb]
\centering
\caption{Rotational periods and peak-to-peak variability amplitudes for J-band variable brown dwarfs.}
\label{mylabel}
\begin{tabular}{lllll}
\hline \hline
\multicolumn{1}{l}{} & 
	 \multicolumn{1}{c}{}& 
		 \multicolumn{1}{c}{Period}& 
			 \multicolumn{1}{c}{$J$-band Amp}  &
			 	\multicolumn{1}{c}{}\\
\multicolumn{1}{l}{Name} & 
	 \multicolumn{1}{c}{Spt}& 
		 \multicolumn{1}{c}{(hr)}& 
			 \multicolumn{1}{c}{(\%)}  &
			 	\multicolumn{1}{c}{Ref}\\								 
\hline
2M0036+18	&L3.5	&$2.7 \pm 0.3$			&$1.22 \pm 0.04$		&	1 \\
W0047		&L6		&$13.2 \pm 0.14$		&$10\pm0.5$			&	2\\
SIMP 0136  	&T2.5	& $2.414 \pm 0.078$    	 & $5$             			&	3, 4\\
SDSS 0423-04  &T0		& $2\pm0.4$      		& $0.8\pm0.08$          	&	5\\ 
2M0559     	&T4.5	& $10\pm3$    			& $0.7\pm0.5$    		&    	3  \\
SDSS 0758  	&T2		& $4.9\pm0.2$      		& $4.8\pm0.2$            	&	3\\
2M0817     	&T6.5	& $2.8 \pm 0.2$   		& $0.6 \pm 0.1$      		& 	3   \\
WISE 1049B 	&T0.5	& $4.87 \pm 0.01$    		& $7\pm0.5$           		& 	6, 7\\
SDSS 1052  	&T0.5	& $3\pm0.5$      		& $2.2\pm0.5$             	&	8\\
DENIS 1058 	&L3		& $4.3 \pm  0.31$  		& $0.843\pm0.098$          	& 	9 \\
2M1126-50     	&L4.5	& $3.2 \pm 0.3$      		& $1.2\pm 0.1$           	& 	3\\
2M1207b     	&L5		& $10.7 \pm 0.8$      		& $1.36\pm 0.23$           	& 	10\\
SIMP 1629  	&T2		& $6.9\pm2.4$      		& $4.3\pm2.4$         		&	3\\
2M1828     	&T5.5	& $5.0 \pm 0.6$     		& $0.9 \pm 0.1$            	&	3\\
PSO-318		&L7.5	&$7.5 \pm 2.5$			&$10 \pm 1$			&	11, 12\\
2M2139+02      	&T1.5	& $7.614 \pm 0.178$  	& 26           			&	4, 13\\
2M2228     	&T6		& $1.369 \pm 0.032$   	& $1.6\pm0.3$          		& 	3, 4 \\
2M2331     	&T5		& $2.9 \pm 0.9$      		& $1.5 \pm 0.2$            	& 	5\\
\hline
\end{tabular}
\\[1.5ex]
{\bf References:} --- (1) \citet{Croll2016},  (2) \citet{Lew2016}, (3) \citet{Radigan2014}, (4) \citet{Yang2016}, (5) \citet{Clarke2008},  (6) \citet{Gillon2013}, (7) \citet{Biller2013}, (8) \citet{Girardin2013}, (9) \citet{Heinze2014}, (10) \citet{Zhou2016}, (11) \citet{Biller2015}, (12) \citet{Allers2016}, (13)\citet{Radigan2012}.
\\[1.5ex]
\label{table:j}
\end{table}

\begin{table}[bh]
\centering
\caption{Rotational periods and peak-to-peak variability amplitudes for \textit{Spitzer} [3.6]$\mu m$ variable brown dwarfs.}
\label{table:spitzer}
\begin{tabular}{lllll}
\hline \hline
\multicolumn{1}{l}{} & 
	 \multicolumn{1}{c}{}& 
		 \multicolumn{1}{c}{Period}& 
			 \multicolumn{1}{c}{$[3.6]~\mu$m Amp}  &
			 	\multicolumn{1}{c}{}\\
\multicolumn{1}{l}{Name} & 
	 \multicolumn{1}{c}{Spt}& 
		 \multicolumn{1}{c}{(hr)}& 
			 \multicolumn{1}{c}{(\%)}  &
			 	\multicolumn{1}{c}{Ref}\\	
\hline
2M0036+18   	&L3.5 	& $2.7 \pm 0.3$ 		& $0.47 \pm 0.05$ 	& 1 \\
2M0050    	&T7		& $1.55 \pm 0.02$ 		& $<0.59 \pm 0.50$ 	& 1   \\
2M0103+19    	&L6		& $2.7 \pm 0.1$   		& $0.56 \pm 0.03$  	&1   \\
2M0107+00    	&L8		& $5 \pm 10 $     		& $1.27 \pm 0.13$   	&1  \\
SIMP 0136  	&T2.5	& $2.414 \pm 0.078$    	& $1.5\pm0.2$          	&2\\
2M0825    	&L7.5	& $7.6 \pm 10$    		& $0.81 \pm 0.08$    	&1\\
WISE0855	&Y1		&$10\pm1$			&$4.5\pm0.5$		&3\\
SDSS1043  	&L9		& $3.8 \pm 0.2$   		& $1.54 \pm 0.15$  	&1   \\
DENIS 1058 	&L3		& $4.1 \pm 0.2$   		& $0.39 \pm 0.04$  	&1  \\
2M1126-50    	&L4.5	& $3.2 \pm 0.3$   		& $0.21 \pm 0.04$   	&1  \\
2M1324    	&T2.5	& $13 \pm 1$      		& $3.05 \pm 0.15$   	&1 \\
WISE1405	&Y0.5	&$8.2\pm0.3$			&$3.6 \pm0.4$		&4\\
2M1507-16    	&L5		& $2.5 \pm 0.1$   		& $0.53 \pm 0.11$  	&1  \\
SDSS1511  	&T2		& $11 \pm 2$      		& $0.67 \pm 0.07$   	&1 \\
SDSS1516  	&T0.5	& $6.7 \pm 10$    		& $2.4 \pm 0.2$  	& 1     \\
2M1615+49   	&L4		& $24 \pm 10 $    		& $0.9 \pm 0.2$   	&1    \\
2M1632    	&L8		& $3.9 \pm 0.2$   		& $0.42 \pm 0.08$ 	&1    \\
2M1721+33    	&L3		& $2.6 \pm 0.1$   		& $0.33\pm 0.07$ 	&1    \\
WISE1738	&Y0		&$6.0\pm0.1$			&$3 \pm 0.1$		&5\\
2M1753    	&L4		& $50 \pm 10$     		& $0.25 \pm 0.5$ 	&1    \\
2M1821+14    	&L4.5	& $4.2 \pm 0.1$   		& $0.54 \pm 0.05$  	&1  \\
HNPegB    	&T2.5	& $18\pm 4$       		& $0.77 \pm 0.15$  	&1  \\
2M2148+40    	&L6		& $19 \pm 4$      		& $1.33 \pm 0.07$    	&1 \\
2M2139+02	&T1.5	&$7.618\pm0.18$		&$11\pm1$		& 2 \\
2M2208+29      &L3		& $3.5 \pm 0.3$   		& $0.69 \pm 0.07$   	&1, 2  \\
2M2228   		&T6		& $1.37 \pm 0.01$ 		& $4.6 \pm 0.2$  	&1  \\

\hline  
\end{tabular}
\\[1.5ex]
{\bf References:} --- (1) \citet{metchev2015a}, (2) \citet{Yang2016}, (3) \citet{Esplin2016}, (4) \citet{Cushing2016},  (5) \citet{Leggett2016}.

\label{table:Spitzer}
\end{table}

Our sample consists of all variable brown dwarfs in the L-T spectral range with published periods and high dispersion NIRSPEC-7 data available in the Keck Archive, as well as three known variable brown dwarfs with measured periods and previously measured $v\sin i$ (WISE 1049B, DENIS 1058 and PSO-318) The full sample is shown in Table \ref{tab:sample} and each object is described briefly below.\\

\textit{2MASS 0036159+182110 --- } The object 2M0036+18 is a magnetically active L3.5 dwarf. Variability was first detected by \citet{Berger2005} in the radio, with a period of $\sim 3~$hr. \added{\citet{Harding2013} detected optical $I-$band variability, confirming the $3~$hr period.} 2M0036+18 was subsequently observed as part of the \textit{Weather on Other Worlds} campaign by \citet{metchev2015a}, who measured a period of $2.7 \pm 0.3~$hr in mid-IR wavelengths. \citet{Croll2016} measure an $J$-band amplitude of $1.22 \pm 0.04 \% $. \citet{Blake2010} have previously measured $v\sin i$ for this object to be $35.12 \pm0.57~$kms$^{-1}$.

\textit{WISE J004701.06+680352.1--- } This very red L6 dwarf was discovered by \citet{Gizis2012}.  \citet{Lew2016} detect $J$-band variability with an amplitude of $10\%$ and a period of $\sim13~$hr. 
They further go on to measure a  $v \sin i=6.7^{+0.7}_{-1.4}$ km s$^{-1}$ and constrain the inclination to $i\sim33^{+5~\circ}_{-8}$. This $v \sin i $ differs from the previously measured value of $4.3\pm2.2~$kms$^{-1}$  by \citet{Gizis2015}. \citet{Gizis2015} assigns an {INT-G} gravity classification to W0047.

\textit{2MASS J0103320+193536 --- } The L6 brown dwarf 2M0103+19 was first monitored by \citet{Enoch2003}, who did not detect $J$-band variability. \textit{Spitzer} observations later revealed mid-IR variability, with an amplitude of $0.47\pm0.05\%$ and a regular $2.7~$hr period \citep{metchev2015a}. This object is given a $\beta$ gravity classification by \citet{Faherty2012} and an {INT-G}  classification by  \citet{Allers2013}.

\textit{2MASS J01075233+0041561 --- } The L8 object 2M0107+00 was observed as part of the  \textit{Weather on Other Worlds} campaign by \citet{metchev2015a}. This is a complex and irregular variable, with an unconstrained period of $5~$hr and an amplitude of $1.27\pm0.12\%$.

\textit{SIMP J0136566+0933473 --- } The variability detection of the T1.5 dwarf SIMPJ0136 by \citet{Artigau2009} was the first highly significant repeatable periodic variability of a brown dwarf at the L/T transition. Long-term monitoring of SIMPJ0136 revealed changes in both amplitude and shape over multiple rotations \citep{Metchev2013}. \citet{Yang2016} constrain the period to $2.414\pm 0.078~$hr and measure a mid-IR amplitude of $1.5\pm0.2\%$.

\textit{SDSS J042348.57-041403.5AB --- } \citet{Enoch2003} reported tentative $K_S$ variability in this T0 binary system. \citet{Clarke2008} monitored SDSS0423-04 in the $J$-band and report low-level variability with a $2~$hr period and a $0.8\pm0.8\%$ amplitude. \citet{Radigan2014} re-observed the binary, finding inconclusive evidence for its variability during a $3.6~$hr observation. 

\textit{WISE J104915.57-531906.1AB --- } WISE 1049B \citep{Luhman2014} is one member of a brown dwarf binary system with spectral types L9 and T0.5 for the A and B components respectively. Variability has been detected in both components \citep{Biller2013,Buenzli2015}. A period of $4.87\pm0.01~$hr has been determined for the B component \citep{Gillon2013}, while a period has not been robustly observed for the A component \citep{Buenzli2015}. \citet{Crossfield2014} report $v\sin i = 26.1\pm0.2~$kms$^{-1}$.

\textit{DENIS 1058.7-1548 --- } Both {\textit{Spitzer}} monitoring and ground-based $J$-band photometry reveal variability in this L3 dwarf \citep{Heinze2014}. DENIS 1058 has a period of $4.3\pm0.31~$hr and amplitudes of $0.39\pm0.04\%$ and $0.843\%$ in the mid-IR and $J$-band respectively. This object is one of five in the sample with both a $J$-band and mid-IR variability detection. DENIS 1058 also has a published $v\sin i = 37.5 \pm 2.5~$kms$^{-1}$  \citep{Basri2000}.

\textit{2MASS J11263991-5003550 --- } 2M1126-50 \citep{Folkes2007} is a peculiar L dwarf with $J-K_S$ colours that are unusually blue for its L4.5 optical or L6.5 NIR spectral type. This target was found to be variable in the $J$-band with a peak-to-peak amplitude of  $1.2\pm0.1 \% $ and a period of $\sim 4~$hr \citep{Radigan2014, Radigan2014a}. \citet{metchev2015a} later constrained the period to $3.2\pm 0.3~$hr via their $0.21\pm0.04\%$ mid-IR variability detection. 

\textit{2MASS J1507476-162738 ---} This L5 object is another irregular variable, showing evidence for spot evolution during the $20~hour$ \textit{Spitzer} observations by \citet{metchev2015a}. The authors determine a period of $2.5\pm0.1~$hr and an amplitude of $0.57 \pm 0.04\%$ for this object. 2M1507-16 has previously measured $v\sin i = 21-30~$kms$^{-1}$ \citep{Bailer-Jones2004, Reiners2008, Blake2010}.

\textit{2MASS J16154255+4953211---} \citet{metchev2015a} detect mid-IR variability in 2M1615+49, and infer a period of $24~$hr and an amplitude of $0.9\pm0.2\%$ from the lightcurve. This object is classified as {VL-G} by \citet{Allers2013} based on FeH and alkali absorption as well as $H$-band shape, however it lacks the deep VO absorption observed in other low-gravity brown dwarfs. \citet{Faherty2016} assigns a $\gamma$ gravity classification. 

\textit{SIMP J16291841+0335380 --- } \citet{Radigan2014} detect $J$-band variability in this T2 dwarf, with an estimated peak-to-peak amplitude of $\sim4.3\%$ and a period of $\sim 6.9 ~$hr. These estimates are uncertain as only the trough of the light curve was caught in the 4 hour observation.

\textit{2MASS J1721039+334415--- } Mid-IR variability was detected in this L3 dwarf by \citet{metchev2015a}, with an inferred period of $2.6 \pm 0.1 ~$hr  and an amplitude of $0.33 \pm 0.07\%$. 

\textit{2MASS J18212815+1414010--- } \citet{metchev2015a} detected mid-IR variability in this L4.5 dwarf, determining a period of $4.2 \pm 0.1~$hr and an amplitude of $0.54\pm0.05\%$. The red near-IR colours and silicate absorption \citep{Cushing2006} of 2M1821+14 indicates an extremely dusty atmosphere, however \citet{Allers2013} and \citet{Gagne2015} find no clear signs of low-gravity. This object has a previously measured $v \sin i = 28.9~$kms$^{-1}$ \citep{Blake2010}. 

\textit{2MASS J1906485+4011068--- } \citet{Gizis2013} detect optical variability in this L1 dwarf using \textit{Kepler}, finding a consistent rotation period of $8.9~$hr with an amplitude of $1.5\%$. \citet{Gizis2013} also report $v \sin i = 11.2 \pm2.2~$kms$^{-1}$ and calculate the inclination, $i > 37^{\circ}$. This is a magnetically active brown dwarf, so the observed variability may be due to magnetic phenomena such as starspots. 

\textit{PSO 318.5 -22 --- }  \citet{Biller2015} detect $J$-band variability in this extremely red exoplanet analog with amplitudes of $7-10\%$ during two consecutive nights of observations. 
PSO-318 has a period of $7.5 \pm 2.5~$hr \citep{Biller2015,Allers2016}. \citet{Allers2016} report a $v\sin i=17.5^{+2.3}_{-2.8}~$kms$^{-1}$  for this object. \citet{Liu2013} classifies this as {VL-G} and \citet{Faherty2016} assigns a $\gamma$ classification.

\textit{2MASS J21392676+0220226 --- } 2M2139+02  is the most variable brown dwarf discovered to date; \citet{Radigan2012} detects variability with $J$-band amplitudes of up to $26 \% $ with a period of $7.721\pm 0.005~$hr. More recently, \citet{Yang2016} monitored 2M2139+02 in 8 separate {\textit{Spitzer}} visits, finding a period of $7.614\pm0.178~$hr, with lower mid-IR amplitudes of $\sim11 \% $. 2M2139+02  is an extreme outlier, exhibiting the highest $J$-band and mid-IR variability amplitudes observed in any brown dwarf to date.

\textit{2MASS 21481628+4003593 --- } \citet{metchev2015a} report mid-IR variability in this L6 dwarf with a period of $19\pm4~$hr and an amplitude of $1.33\pm0.07\%$.

\textit{2MASS 2208136+292121 --- } \citet{metchev2015a} observed variability in this L3 brown dwarf. A period of $3.5\pm 0.2~$hr and an amplitude of $0.62\%$ were determined from the lightcurve. 2M2208+40 has been assigned $\gamma$ and VL-G classifications \citep{Cruz2009,Allers2013}.


\subsection{Low-Gravity Brown Dwarfs}
As discussed in Section \ref{sec:thesample}, the brown dwarfs \textit{2M0103+19, 2M1615+49, 2M2208+29, PSO-318} and \textit{W0047} show signs of low-gravity. Low-gravity is indicative of both a lower mass and a larger radius, which in turn is suggestive of a young brown dwarf that has not yet contracted to reach its equilibrium radius. This subsample provides valuable information on the effects of gravity and youth on variability properties.
\citet{metchev2015a} note a tentative correlation between low-gravity and high-amplitude mid-IR variability amplitudes. This correlation is further supported by a number of high-amplitude $J$-band detections in low-gravity objects \citep{Biller2015, Lew2016}. This is unexpected because atmospheric models typically require very thick clouds \citep{Madhusudhan2011} and initial variability studies suggest that objects with patchy clouds in the process of breaking up tend to have the highest variability amplitudes \citep{Radigan2014}. Evidently, low-gravity objects can exhibit very different atmospheric properties to field brown dwarfs, and they are denoted by a black inset in all plots in this paper.


\section{Data and Observations}
\label{sec:data}
We obtained high dispersion NIRSPEC spectra for our targets from the Keck Observatory Archive. NIRSPEC is a near-infrared echelle spectrograph on the Keck II $10~m$ telescope on Mauna Kea, Hawaii. The NIRSPEC detector is a $1024 \times 1024$ pixel ALADDIN InSb array. Observations were carried out using the NIRSPEC-7 ($1.839 - 2.630~ \mu m$) passband in echelle mode using the 3 pixel slit ($0.432''$) an echelle angle of $62^{\circ}.67-63^{\circ}.00$ and a grating angle of $35^{\circ}.46-35^{\circ}.52$. Observations of targets were gathered in nod pairs, allowing for the removal  of sky emission lines through the subtraction of two consecutive images. Arclamps were observed for wavelength calibration. $5-10$ flat field and dark images were taken for each target to account for variations in sensitivity and dark current on the detector. Details of the observations are given in Table \ref{tab:NIRSPEC}.

\begin{table*}[tb]
\centering

\caption{NIRPSEC-7 high dispersion observing information. All data were taken from the Keck archive.}
\label{tab:NIRSPEC}
\begin{tabular*}{0.82\textwidth}{lllllllll}
\hline \hline
\multicolumn{1}{l}{} & 
	 \multicolumn{1}{c}{}& 
		 \multicolumn{1}{c}{}& 
			 \multicolumn{1}{c}{Echelle}  &
			 	\multicolumn{1}{c}{Cross Disp}&
					\multicolumn{1}{c}{Exp Time}&
			 			\multicolumn{1}{c}{}&
			 				\multicolumn{1}{c}{}&
								 \multicolumn{1}{c}{}\\
\multicolumn{1}{c}{Name} & 
	 \multicolumn{1}{c}{UT Date}& 
		 \multicolumn{1}{c}{Slit Name}& 
			 \multicolumn{1}{c}{(deg)}  &
			 	\multicolumn{1}{c}{(deg)}&
			 		\multicolumn{1}{c}{(s)}&
						\multicolumn{1}{c}{Airmass}&
			 				\multicolumn{1}{c}{S/N}&
							 	\multicolumn{1}{c}{Prog ID}\\
\hline
2M0036+18   & 2011-09-10 & $0.432 \times12$ & 63.00      & 35.52      & $2\times450$                 & 1.006		& 28  & U049NS  \\
W0047    & 2013-09-17 & $0.432\times12$ & 62.97              & 35.51      & $2\times1200$                & 1.507& 24  & U055NS  \\
2M0103+19   & 2014-07-19 & $0.432\times24 $  & 62.68     & 35.44      & $2\times300 $                & 1.209 & 10  & N160NS  \\
2M0107+00   & 2011-09-07 & $0.432\times12   $& 63.00      & 35.46      & $2\times1500$                & 1.070 & 24  & U049NS  \\
SIMP1036 & 2011-09-10 & $0.432\times12   $& $63.00$       & 35.52      & $2\times600  $               & 1.061 & 17  & U049NS  \\
SDSS0423-04 & 2004-03-08 &$0.432\times12   $& 62.65     & 35.51      & $2\times1200$                & 1.342 & 21  & C13NS   \\
2M1126-50   & 2014-01-20 & $0.432\times12   $& 63.02       & 35.53      & $2\times600  $               & 2.892 & 15  & U055NS  \\
2M1507-16   & 2011-06-10 & $0.432\times12   $& 63.00       & 35.53      & $2\times600      $           & 1.288 & 19  & U038NS  \\
2M1615+49   & 2011-09-10 & $0.432\times12   $   & 63.00   & 35.52      & $2\times900    $             & 1.491 & 21  & U049NS  \\
SIMP 1629 & 2011-09-07 & $0.432\times12   $  & 63.00             & 35.46      & $2\times1200  $              & 1.119 & 24  & U049NS  \\
2M1721+33   & 2011-09-07 & $0.432\times12   $   & 63.00            & 35.46      & $2\times1000 $               & 1.254 & 27  & U049NS  \\
2M1821+14  & 2006-07-30 & $0.432\times12   $   & 62.67         & 35.51      & $2\times600   $              & 1.075 & 26  & N050NS  \\
2M1906+40   & 2011-09-10 & $0.432\times12   $   & 63.00           & 35.52      & $2\times120    $             & 1.631 & 10  & U049NS  \\
2M2139+02    & 2011-09-07 &$0.432\times12   $   & 63.00             & 35.46      & $2\times1200    $            & 1.048 & 23  & U049NS  \\
2M2148+40   & 2006-12-32 & $0.432\times12   $   & 62.68         & 35.52      & $2\times750      $           & 1.451 & 22  & N044NS  \\
2M2208+29  & 2011-09-10 & $0.432\times12   $ & 63.00             & 35.52      & $2\times1500        $        & 1.054 & 27  & U049NS \\
\hline
\end{tabular*}
\\[1.5ex]
\label{table:results}
\end{table*}

\section{Data Reduction Methods}
\label{sec:methods}
Data were reduced using a modified version of the REDSPEC reduction package to spatially and spectrally rectify each exposure. The KECK/Nirspec Echelle Arc Lamp Tool was used to identify the wavelengths of lines in our arc lamp spectrum. 
We focus our analysis on order 33 since this part of the spectrum contains a good blend of sky lines and brown dwarf lines, allowing for an accurate fit. Order 33 is also commonly used in the literature for NIRSPEC high dispersion N-7 spectra \citep{Blake2010, Gizis2013}. We additionally reduce orders 32 and 38, which again contain a sufficient amount of sky and brown dwarf lines, to check for consistency.
 After nod-subtracting pairs of exposures, we create a spatial profile which is the median intensity across all wavelengths at each position along the slit. To remove any residual sky emission lines from our nod-subtracted pairs we identify pixels in the spatial profile that do not contain significant source flux. We use Poisson statistics to determine the noise per pixel at each wavelength. We extract the flux within an aperture in each nod-subtracted image to produce two spectra of our source.
  The extracted spectra are combined using a robust weighted mean with the xcombspec procedure from the SpeXtool package \citep{Cushing2004}.

\subsection{Determining Rotational Velocities}
We use the approach outlined in \citet{Allers2016} to determine the rotational velocities of our objects. We employ forward modelling to simultaneously fit the wavelength solution of our spectrum, the rotational and radial velocities, the scaling of telluric line depths, and the FWHM of the instrumental line spread function (LSF). \added{We use the BT-Settl model atmospheres \citep{Allard2012} as the intrinsic spectrum for each of our objects. Further details can be found in \cite{Allers2016}.} 
 In total, the forward model has nine free parameters: the $T_\mathrm{eff}$ and $log(g)$ of the atmosphere model, the $v_r$ and $v \sin i$ of the brown dwarf, $\tau$ for the telluric spectrum, the LSF FWHM, and the wavelengths of the first, middle and last pixels. The forward model is compared to our observed spectrum, and the parameters used to create the forward model are adjusted to achieve the best fit.

\begin{figure}[tb]
	\centering{
		\includegraphics[scale=0.45]{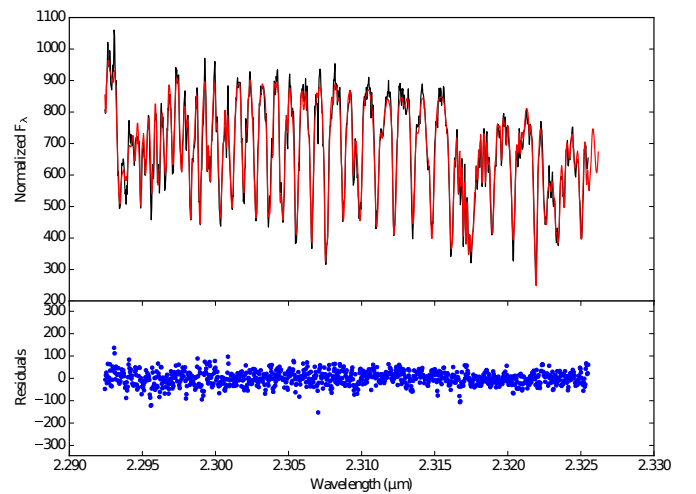}}
	\caption{ The observed spectrum of 2M1507-16 (black) compared to our forward model with best fit parameters (red). }
	\label{fig:modelfit}
\end{figure}

To determine the best fit parameters of our forward model as well as their marginalised distributions we use a Markov Chain Monte Carlo (MCMC) approach. This involves creating forward models that allow for a continuous distribution  of $T_{\mathrm{eff}}$ and $log(g)$ by linearly interpolating between atmosphere grid models. We employ the DREAM(ZS) algorithm \citep{terBraak2008}, which uses an adaptive stepper, updating model parameters based on chain histories. An example of our best fit model for 2M1507-16 order 33 is shown in Figure \ref{fig:modelfit}. 
\added{Table \ref{table:results} shows the resulting rotational velocities, radial velocities, effective temperatures and surface gravities calculated using order 33. We also find that orders 32 and 38 are consistent with the results obtained from order 33.}

As discussed in Section \ref{sec:thesample}, 5 of the objects in our sample have previous measurements of $v\sin i$. Our value of $35.91^{+0.8}_{-0.8}~$kms$^{-1}$  for 2M0036+18 is consistent with the $v \sin i$ measured by \citet{Blake2010}. Literature $v \sin i$ measurements  for 2M1507-16 have ranged from $21-30~$kms$^{-1}$ \citep{Bailer-Jones2004, Reiners2008, Blake2010} and we find that our measurement of $19.21^{+0.53}_{-0.53}~$kms$^{-1}$ is consistent with the \citet{Blake2010} measurement. Our measurement of $30.61^{+0.69}_{-0.69}~$kms$^{-1}$  for 2M1821+14 is slightly larger than the \citet{Blake2010} measurement of $28.9\pm0.16~$kms$^{-1}$, but is in agreement within $2\sigma$. Our $v\sin i$ measurement for 2M1906+40 is slightly larger than the \citet{Gizis2013} measurement, but is again consistent within $2\sigma$. Finally, our measurement of $9.6^{+0.49}_{-0.49}~$kms$^{-1}$ for W0047 is higher than both previous measurements by \citet{Gizis2015} and \citet{Lew2016}. 
\added{The model atmosphere for W0047 used by \citet{Gizis2015} has $T_{\mathrm{eff}}=2300$ and $\log(g)=5.5$, while evolutionary models predict $T_{\mathrm{eff}}=1270$ and $\log(g)=4.5$ \citep{Gizis2015}. Our model (with $T_{\mathrm{eff}}=1670$ and $\log(g)=5.2$) is in better agreement with the evolutionary model results. With higher effective temperature and surface gravity, the atmospheric model used by \citet{Gizis2015} will include more pressure broadening, and thus result in a lower value of $v\sin i$. \citet{Lew2016} do not provide details on the atmospheric model used. Again, the consistency between orders 32, 33 and 38 further supports our results.}

\begin{table*}[tp]
\setlength{\tabcolsep}{10pt}
\centering
\caption{Rotational velocities, radial velocities, effective temperatures and gravities calculated in this study. The results presented here are for order 33. The last two columns show our estimated radii and the resulting angle of inclination calculated for each object. }
\begin{tabular*}{0.95\textwidth}{cccccccc}
\hline \hline
\multicolumn{1}{c}{Name} & 
	 \multicolumn{1}{c}{$v\sin i $}& 
	 	 \multicolumn{1}{c}{RV}& 
	 		  \multicolumn{1}{c}{$T_{\mathrm{eff}}$}& 
	  			  \multicolumn{1}{c}{$\log(g) $}& 
					 \multicolumn{1}{c}{Radius}& 
						 \multicolumn{1}{c}{Inclination}  \\
\multicolumn{1}{c}{} & 
	 \multicolumn{1}{c}{$	(kms^{-1})$}& 
	 	 \multicolumn{1}{c}{$	(kms^{-1})$}& 
		 	 \multicolumn{1}{c}{K}& 
			 	 \multicolumn{1}{c}{dex}& 
					 \multicolumn{1}{c}{($R_{Jup}$)	}& 
						 \multicolumn{1}{c}{($^{\circ}$)}  \\
\hline
2M0036+18  	& $36.0\pm0.2$   		&$20.9\pm0.14$		&$1909\pm6$			&$5.22\pm0.02$   	 	&$0.94-1.08^{\mathrm{a}}$            	&  $51 \pm 9$      	\\
W0047		& $9.8\pm0.3 $			&$-19.8_{-0.2}^{+0.1}$	&$1666\pm2$			&$5.16_{-0.3}^{+0.2}$	&$1.26-1.34^{\mathrm{a}}$			& $53 \pm 3$		\\ 
2M0103+19	&$40.0_{-4.7}^{+3.7}$	&$12.4_{-4.0}^{+3.8}$	&$1880_{-110}^{+200}$	&$4.0_{-0.4}^{+0.6}$		&$1.21-1.47^{\mathrm{a}}$			& $40 \pm 8$		\\
2M0107+00   	& $19.4\pm0.8$    		&$8.2\pm0.5$			&$1450_{-20}^{+70}	$	&$4.7_{-0.1}^{+0.4}$		& $0.87-1.09^{\mathrm{a}}$          	&  $56 \pm 17$     	\\
SIMP 0136 	& $52.8_{-1.0}^{+1.1}  $    &$12.3\pm0.8$			&$1290\pm10$			&$5.45_{-0.04}^{+0.03}$   & $0.8-1.2$           		&  $80 \pm12$       	\\ 
SDSS 0423-04	&$68.0\pm0.9$			&$30.5\pm0.6$			&$1460\pm10$			&$5.27_{-0.04}^{+0.5}$	&$0.8-1.2$			&$79_{-16}^{+11}$	\\
WISE 1049B$^{\mathrm{b}}$	& $26.1\pm0.2$     &				&					&					&$ 0.95-1.09^{\mathrm{a}}$  		&$83_{-8}^{+7}$	\\
DENIS 1058$^{\mathrm{c}}$ 	&  $37.5\pm2.5$   &				&					&					&$ 0.93-1.07^{\mathrm{a}}$  		&$90_{-2}$			\\	
2M1126-50   	&$22.8_{-2.4}^{+1.6}  $ 	&$49.3\pm1.1$			&$1270_{-20}^{+60}$	&$3.7_{-0.1}^{+0.5}$		& $0.8-1.2$        		& $35 \pm 7$  		\\
2M1507-16   	& $19.1\pm0.5$		&$-39.2_{-0.4}^{+0.3}$	&$1750\pm6$			&$5.45\pm0.04$		& $0.9-1.08^{\mathrm{a}}$            	&  $23 \pm2$		\\
2M1615+49   	& $9.5_{-1.2}^{+1.3} 	$	&$-21.3\pm0.5$		&$1624_{-48}^{+8}$		&$4.53_{-0.1}^{+0.08}$	& $1.1-1.4$        		&  $86^{+4}_{-10}$  		\\
SIMP 1629 	&  $19.7_{-0.8}^{+0.7}$	&$7.7\pm0.5$			&$1277\pm7$			&$5.29\pm0.03$	      	& $ 0.8-1.2$        		& $82^{+8}_{-13}$     \\
2M1721+33   	& $21.5\pm0.3  $		&$-102.8\pm0.2$		&$1656\pm2$			&$4.77\pm0.02$		& $ 0.8-1.2$         		&  $27\pm 4$     	\\
2M1821+14   	& $30.7\pm0.2   $ 		&$11.0\pm0.1$			&$1766\pm1$			&$4.89\pm0.01$		& $ 0.8-1.2$       		&  $61 \pm17$      	\\
2M1906+40   	& $15.2\pm0.5   $  		&$-22.8_{-0.2}^{+0.3}$	&$1999_{-5}^{+3}$		&$5.30\pm0.04$  		& $0.8-1.2$            		&$82_{-12}^{+8}$	\\
PSO-318$^{\mathrm{d}}$		&  $17.5^{+2.3}_{-2.8}$ 	&$6.0_{-1.1}^{+0.8}$		&$1325^{+330}_{-12}$	&$3.7^{+1.1}_{0.1}$		&$1.38-1.44^{\mathrm{a}}$			& $61 \pm 17$		\\
2M2139+02   	&  $18.7\pm0.3   $		&$-25.1\pm0.3$		&$1333\pm5$			&$5.37\pm0.02$		& $0.8-1.12^{\mathrm{a}}$         		 &  $90_{-1}$       	\\
2M2148+40	& $9.2_{-0.3}^{+0.4}$	&$-14.3\pm0.1$		&$1774\pm1$			&$5.00_{-0.02}^{0.01}$	& $0.89-1.09^{\mathrm{a}}$			&$88^{+2}_{-8}$		\\
2M2208+29   	&  $40.6_{-1.4}^{+1.3}   $  &$-15.7^{+0.8}_{-0.9}$	&$1707^{+10}_{-9}$ 		&$4.27\pm0.11$       		&$1.21-1.61^{\mathrm{a}}$            	& $55\pm 10$    		\\ 

\hline
\end{tabular*}
\\[1.5ex]
\raggedright{
$^{\mathrm{a}}$ Radii are taken from \citet{Filippazzo2015}\\
$^{\mathrm{b}}$ $v\sin i$ measurement taken from \citet{Crossfield2014}\\
$^{\mathrm{c}}$ $v\sin i$ measurement taken from \citet{Basri2000}\\
$^{\mathrm{d}}$ Measurements taken from \citet{Allers2016}

}
\label{table:results}
\end{table*}

\subsection{Calculating Inclination Angles}
We assume that the brown dwarf rotates as a rigid sphere. However, this is not strictly true. The rotational period of Jupiter, as measured by magnetic fields originating in the core is $9^h50^m30^s$, whereas the period measured using features rotating along the equator is $9^h55^m40^3$, a difference of only $5~$minutes. Since rotational periods as measured from photometric variability in general have much larger uncertainties, the rigid body assumption is reasonable for our analysis. Thus, the equatorial rotation velocity, $v$, is given by $v=2\pi R / P$, where $R$ is the radius of the brown dwarf and $P$ is its rotation period. With our measured values of $v \sin i$ in hand, an assumption of radius and a measurement of the rotation period allow us to determine the angle of inclination, $i$.
\citet{Filippazzo2015} provide radius estimates from evolutionary models for 11/19 of our targets (starred in Table \ref{table:results}). We use reasonable radius estimates for the remaining field brown dwarfs. At field brown dwarf ages, the radii are independent of mass due to electron degeneracy \citep{Burrows2001} and approach the radius of Jupiter. Therefore, the field brown dwarf targets are assumed to have a radius of $0.8-1.2~M_{Jup}$. 2M1615+49 is the only young brown dwarf with no radius estimate. Since it has not been associated with any moving group \citep{Faherty2016}, we have no age constraint on this object. We assume a radius of $1.1-1.7~M_{Jup}$, similar to other VL-G objects in the sample. 

\added{Monte Carlo analysis is used to determine the inclination, $i$ for each target, using uniformly distributed radii and gaussian distributions for the $v\sin i$ and period values. The inclination and error are calculated as the mean and standard deviation of the resulting distribution of $i$. 
Table \ref{table:results} shows the rotational velocities calculated for our sample, as well as the inclination angles determined based on our estimated radii. As stated earlier, we focus our analysis on order 33. However using a weighted-average of $v \sin i$ values obtained from orders 32, 33 and 38 yields consistent inclination angles.}

\section{Results and Discussion}
\label{sec:results}


\subsection{Effects of Inclination on Variability Amplitude}

\begin{figure}[]
	\centering
		\includegraphics[scale=0.51]{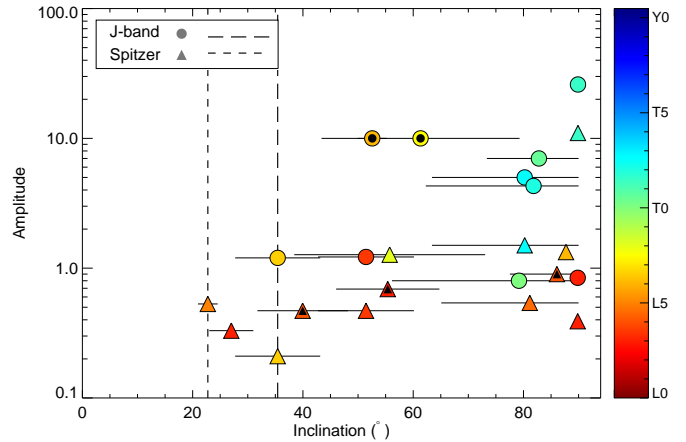}
	\caption{Variability amplitude plotted against inclination angle for our sample. Circles denote $J$-band detections while triangles denote \textit{Spitzer} 3.6 $\mu m$ detections. The colour scale represents spectral type and young objects are denoted by a black inset. Dashed lines represent the minimum inclination angle for each band.  }
	\label{fig:inc_amp}
\end{figure}

Figure \ref{fig:inc_amp} shows variability amplitude plotted against the angle of inclination. We note a number of interesting trends in the $J$-band and \textit{Spitzer} variable brown dwarfs. 

Firstly, the highest amplitude $J$-band variable objects are either L/T transition brown dwarfs  or young, red brown dwarfs. The highest \textit{Spitzer} and $J$-band amplitudes are both for the L/T transition brown dwarf, {2M2139+02 }. The \textit{Spitzer} amplitudes for young brown dwarfs are slightly enhanced, but only relative to their own spectral type and not the entire \textit{Spitzer} sample.

\begin{figure}[]
	\centering
		\includegraphics[scale=0.51]{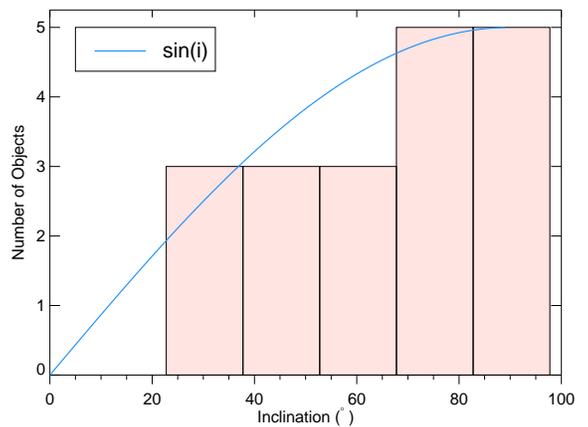}
	\caption{Histogram showing the distribution of inclinations in our sample. The probability distribution of randomly oriented objects, $P(i)\sim \sin (i)$. This distribution fits the calculated distribution quite well }
	\label{fig:inc_dist}
\end{figure}

Secondly, while it is clear that each brown dwarf has its own intrinsic amplitude, the inclination angle affects the observed amplitude for both bands.
Figure \ref{fig:inc_amp} shows that there are no mid-IR variability detections at inclination angles $<20^{\circ}$ and no $J$-band detections at inclination angles $<35^{\circ}$. \added{ For a sample of objects with random orientation, the probability distribution of the inclination angles, $P(i) \sim \sin i$ \citep{Jackson2010}. Thus, the  overall observed distribution is fairly consistent with the distribution expected for brown dwarfs that are randomly oriented in space (Figure \ref{fig:inc_dist}). This means that although our sample is small, it is representative of the brown dwarf population with regard to inclination. } Excluding the young objects, we find relatively low amplitudes at inclination angles $20-60^{\circ}$. At inclinations close to $90^{\circ}$ we observe the highest variability amplitudes in both bands.
This makes sense as the brown dwarf is nearly equator-on, allowing us to observe the full variability amplitude. An atmospheric feature observed on a low inclination object will appear smaller due to projection effects.

The $J$-band amplitudes appear to be more affected by inclination than the \textit{Spitzer} amplitudes. The highest $J$-band variable objects appear at high inclinations, whereas a \textit{Spitzer} brown dwarf viewed equator-on displays similar amplitudes to those observed at inclinations as low as $\sim 20^{\circ}$. This may be explained by considering the pressures probed by each band. \citet{Biller2013}, \citet{ Buenzli2012} and \citet{Yang2016} determined the pressure level probed at optical depth $\tau = 2/3$ as a function of wavelength for various models, finding that the $J$-band probes a discrete range of pressures deep in the atmosphere. On the other hand, the \textit{Spitzer} [$3.6~\mu$m] band probes a broader range of pressures, that extend higher up in the photosphere. For the deep layers probed by the $J$-band, the flux will be strongly attenuated for the low-inclination objects due to an increased path length through the atmosphere. The effect is not observed as strongly for \textit{Spitzer} detections because more of the flux originates from near the top of the photosphere. Thus, we see $J$-band amplitudes decrease strongly with decreasing inclination.

\begin{figure}[]
	\centering
		\includegraphics[scale=0.5]{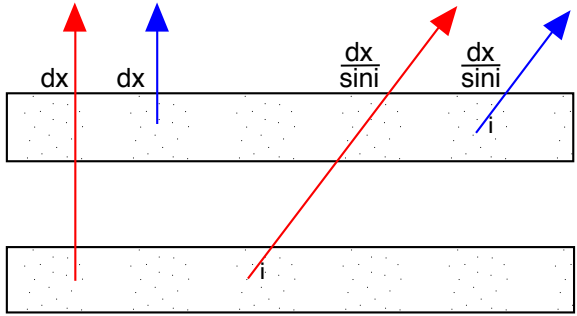}
	\caption{The inclination angle $i$ affects the atmospheric path length travelled from flux originating from a certain depth. In this diagram above, the bottom shaded area corresponds to the depth at which most of the $J-$band flux originates from when an object is viewed equator-on. The top shaded area corresponds to the depth at which most of the \textit{Spitzer} $3.6/mu$m flux comes from for an equator-on object. At $i=90^{\circ}$, the flux is attenuated by $\kappa \mathrm{dx}$ where $\kappa$ is the attenuation coefficient and $\mathrm{dx}$ is the distance to the top of the atmosphere.  At $i<90^{\circ}$ this flux is more strongly attenuated due to a longer atmospheric path length.  }
	\label{fig:inkscape}
\end{figure}

\begin{figure}[]
	\centering
		\includegraphics[scale=0.51]{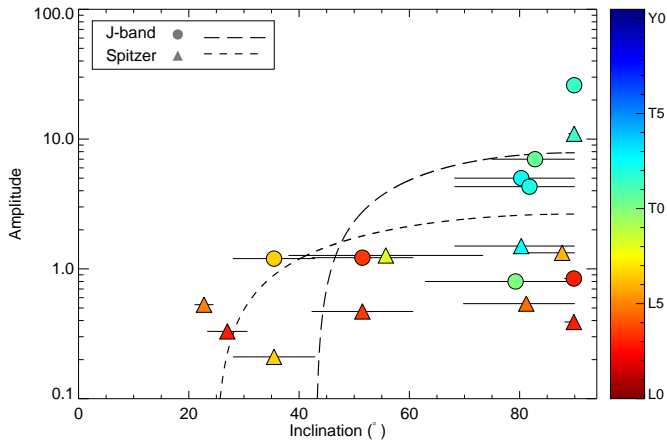}
	\caption{Variability amplitude plotted against inclination angle for \textit{Spitzer} 3.6 $\mu m$ (triangles) and J-band (circles) field objects in our sample. The colour scale represents spectral type and young objects are denoted by a black inset. Best fit functions of Equation \ref{eq:toymodel} are plotted as grey dashed lines. }
	\label{fig:inc_amp_model}
\end{figure}
\begin{table}[]
\centering
\caption{Best-fit parameters for Equation \ref{eq:toymodel}. Best-fit functions for both bands are plotted in Figure \ref{fig:inc_amp_model}.}
\label{model_params}
\begin{tabular}{lll}
\hline \hline
            & $J$-band       & \textit{Spitzer} {[}$3.6~\mu$m{]} \\
           \hline
$A_0$       & $14.69\pm0.11$ & $3.20\pm0.06$            \\
$\kappa \mathrm{dx}$ & $6.85\pm0.07$ & $0.56\pm0.03$     \\
\hline \hline      
\end{tabular}
\end{table}

We use a toy model to investigate the effects of inclination on the observed variability amplitude. Our model has two terms:
\begin{equation}
\label{eq:toymodel}
	A=A_0\mathrm{sin}i -\kappa  \frac{\mathrm{dx}}{\mathrm{sin}i}
\end{equation}
where $A$ is the observed amplitude \replaced{$A_0$ is the intrinsic amplitude $\kappa$ is the attenuation coefficient and $\frac{dx}{\mathrm{sin}i} $ is the atmospheric path length.} {and $A_0$ is the amplitude that would be observed if there were no atmospheric attenuation of the flux. $\kappa$ is the factor by which the flux is attenuated as it passes through the atmosphere and $\frac{dx}{\mathrm{sin}i} $ is the atmospheric path length.}
The first term is a projection effect, which causes the observed area of a spot to decrease as the brown dwarf approaches lower inclinations. The second term represents the attenuation of the flux as it passes through the brown dwarf atmosphere. \added{ Figure \ref{fig:inkscape} shows how decreasing the inclination angle increases the atmospheric path length. }
From the models discussed above, we expect that the $J$-band path lengths are larger than the \textit{Spitzer} path lengths. We fit the function for both bands, assuming that all objects have the same intrinsic amplitude. We  consider only the field brown dwarfs since young objects will have very different atmospheric structures. The best fit functions are shown in Figure \ref{fig:inc_amp_model}. The model fits the data reasonably well, displaying the earlier drop-off of the $J$-band amplitudes compared to the \textit{Spitzer} amplitudes due to a much larger $J$-band $\kappa  {\mathrm{dx}}/{\mathrm{sin}i}$ term.
\deleted{Using a power law for the extinction, $\kappa \sim \lambda^{- \alpha}$, we find that $-2<\alpha < 0$ produces relative path lengths $\mathrm{dx}_{3.6\mu \mathrm{m}} / \mathrm{dx}_{J}$ of $0.05 - 0.45$. This is in agreement with the relative depths calculated for L5, T2 and T6 brown dwarfs by {Yang2016}.}
\added{We estimate the brown dwarf atmospheric extinction as a power law: $\kappa \sim \lambda^{- \alpha}$, where $\alpha=1.7$  \citep{Bertoldi1999}. While this is an empirical law based on extinction by the interstellar medium, dust grains found in the atmospheres of brown dwarfs may be similar in size and thus produce similar results \citep{Looper2010, Marocco2014}. Thus, by estimating the extinction coefficient, we can estimate the relative path lengths travelled by the flux in each band. We find that $\mathrm{dx}_{3.6\mu \mathrm{m}} / \mathrm{dx}_{J}=0.40$. 
\citet{Yang2016} calculate the pressure levels probed at optical depth $\tau = 2/3$ as a function of wavelength for models with a range of spectral types. For all spectral types investigated, they find that the $J$-band probes a discrete range of pressures deep in the atmosphere, while the pressures probed by the \textit{Spitzer} [$3.6~\mu$m] extend higher in the atmosphere.
The relative pressures found in this study for L5, T2 and T6 brown dwarfs were $\mathrm{P}_{3.6\mu \mathrm{m}} / \mathrm{P}_{J}=0.39,0.05$ and $0.05$ respectively. If we assume that the depth increases monotonically with pressure then our value of $\mathrm{dx}_{3.6\mu \mathrm{m}} / \mathrm{dx}_{J}$ is consistent with that of the L5 brown dwarf computed by \citet{Yang2016}. Of course this is a highly simplistic model with some limitations. Firstly, it does not take into account spectral types or different intrinsic variabilities. Secondly, since the majority of $J-$band variability detections are from ground-based surveys, they are not sensitive to the lower amplitudes detected by \textit{Spitzer} in the mid-IR. Thirdly, the model fits are strongly influenced by the absence of detections at low-amplitudes, however this may be due to the underlying inclination distribution and not because their variability amplitudes are below detection limits.
}

\subsection{Relation between Period and Variability Amplitude}
\begin{figure}[tb]
	\centering
		\includegraphics[scale=0.51]{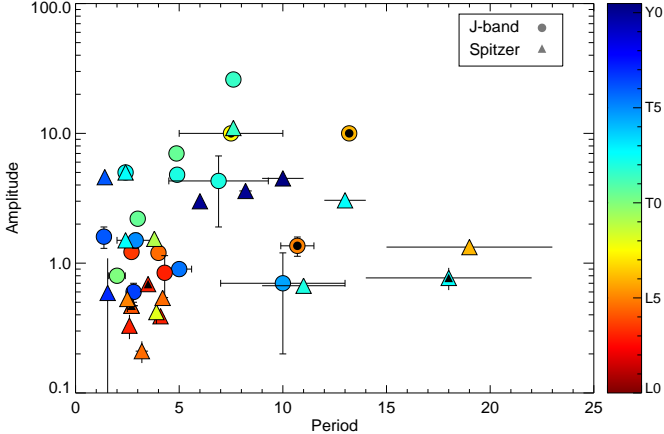}
	\caption{Variability amplitude plotted against period for \textit{Spitzer} 3.6 $\mu m$ (triangles) and J-band (circles) variability detections. The colour bar represents the spectral type of each object and young objects are denoted by a black inset. Objects with unconstrained periods from were not included. \added{Data and literature references are shown in Tables \ref{table:j} and \ref{table:spitzer}.}}
	\label{fig:per_amp_full}
\end{figure}

Figure \ref{fig:per_amp_full} shows the variability amplitude plotted against rotation period for \textit{Spitzer} and $J$-band variable L, T and Y spectral type objects with published periods from the literature \added{ (shown in Tables \ref{table:j} and \ref{table:spitzer}). 
The mid-IR \textit{Spitzer} detections are extremely robust due to the high photometric precision achievable from space. Additionally, these observations are typically longer than ground-based observations -- for example, \citet{metchev2015a} employ $\sim20~$hr observations in their survey. This results in extremely accurate period measurements for \textit{Spitzer} monitored objects.  In contrast, the $J-$band detections come from a variety of ground-based and space-based HST surveys. The ground-based searches do not reach the same photometric precision as space-based searches and thus are limited to higher amplitudes. $J-$band monitoring observations are shorter than \textit{Spitzer} observations and thus have larger period uncertainties. For both samples, we only take objects whose periods are constrained.
} 

The $J$-band and \textit{Spitzer} data display notably different period and variability amplitude properties. Ground-based $J$-band detections have lower photometric precision, so in general $J$-band detections are limited to larger amplitudes. It is clear that mid-IR variability is intrinsically lower than near-IR variability however, as high amplitude variability would certainly have been detected with \textit{Spitzer}. Ground-based observations are only sensitive to shorter periods ($>15~$hr), so the longer period variable brown dwarfs have been detected with \textit{Spitzer}.

\begin{figure}[tb]
	\centering
		\includegraphics[scale=0.51]{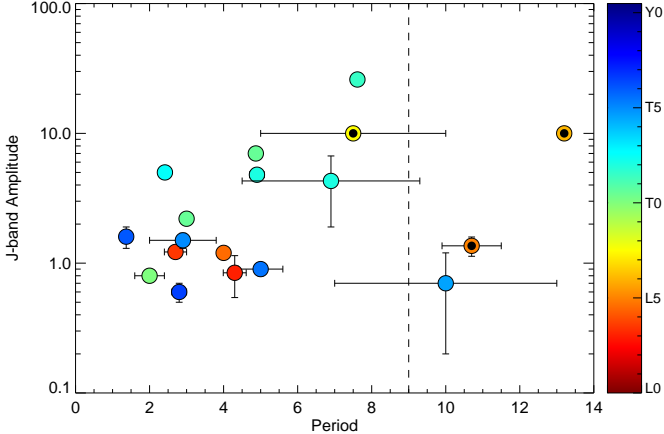}
	\caption{Variability amplitude plotted against period for J-band variability detections. The colour bar shows spectral type and young objects are denoted by a black inset. The dashed line shows the cut-off point of the period range for which the Rhines scale appears to have an effect on variability amplitude. \added{For rotation periods $<9~$hr, we find a tentative correlation between variability amplitude and period with a $p-$value $=6.7\%$.}}
	\label{fig:per_amp_j}
\end{figure}

Figure \ref{fig:per_amp_j} shows the variability amplitude plotted against rotation period for all $J$-band variable objects with published periods (shown in Table \ref{table:j}). Measured periods are $<15~$hr, since most $J$-band detections are ground-based, and thus are sensitive to this range of periods. The highest amplitudes are  L/T transition spectral types, as reported by \citet{Radigan2014}.\added{The young, low-gravity L-type objects {W0047}, {PSO-318} and HNPegb display higher variability amplitudes than other L dwarfs, supporting a tentative correlation between low-gravity and high-amplitude variability reported by \citet{metchev2015a}.} Additionally, for periods $\sim7-9~$hr, there seems to be an overall increase in $J$-band variability amplitude with longer periods.

\replaced{We calculate the significance of this result by calculating the Pearson correlation coefficient using IDLs \textit{correlate.pro}. The correlation coefficient is given by:
where $X$ and $Y$ are the colour anomalies and inclinations and $n$ is the number of objects.
The Pearson correlation coefficient is a measure of linear correlation. To calculate the significance of this correlation we calculate the test-statistic, given by
which follows a t-distribution. Using this distribution we calculate the probability of obtaining a value $\geq t$ assuming that no correlation exists. This is known as the $p$-value.}
{We calculate the significance of this result by calculating Kendall's $\tau$ using IDLs \textit{r\_correlate.pro}. Kendall's $\tau$ is a nonparametric measure of correlation based on the relative ordering of the rank of each value in the dataset \citep{Press1987}. To define $\tau$, we start with $N$ data points $(x_i,y_i)$, and consider all $\frac{1}{2} N (N-1)$ pairs of data points. A pair is \textit{concordant} if the relative ordering of the ranks of $(x_i,x_j)$ is the same as the relative ordering of the ranks of $(y_i,y_j)$. A pair is \textit{discordant} if the relative ordering of   $(x_i,x_j)$ differs from the ordering of the  $(y_i,y_j)$ ranks.  When the relative $(x_i,x_j)$ ranks are the same, we call the pair an "extra-$y$" pair. Similarly, when  relative $(x_i,x_j)$ ranks are the same, we get an "extra-$x$" pair. Kendall's $\tau$ is then calculated using the equation:}
\begin{equation}
	\tau = \frac{\mathrm{C} - \mathrm{D}} { \sqrt{\mathrm{C}+			 \mathrm{D}+\mathrm{extra-}y}  \sqrt{\mathrm{C}+\mathrm{D}+\mathrm{extra-}x}  }
\end{equation}
\added{where C and D are the number of concordant and discordant pairs respectively. In the null hypothesis of no association between $x$ and $y$, $\tau$ is normally distributed with zero expectation value and a variance of}
\begin{equation}
	\mathrm{Var}(\tau) = \frac{N+10}{9N(N-1)}
\end{equation}
\added{Using this distribution we calculate the probability of obtaining a value $\geq t$ assuming that no correlation exists. This is known as the $p$-value.}

Calculating the Kendall's $\tau$ rank correlation coefficient  and $p$-value, we find that the relation between $J$-band variability amplitude and rotational period (for periods $<9~$hr) is significant with a \replaced{$p$-value $<1\%$}{$p$-value $=6.7\%$. In contrast, including all periods, the correlation between period and amplitude is not significant, with a $p$-value$=17\%$.}
\replaced{This increase in variability amplitude with increasing rotation period}{This tentative correlation between variability amplitude and rotation period for periods $<9~$hr} may be explained by consideration of the Rhines length \citep{Rhines1970}. Organised jet features in the atmospheres of the giant Solar System planets generally scale in size with the Rhines length. This also represents the maximum attainable size that a coherent atmospheric structure can grow to before being destroyed by such zonal jets. The Rhines length is given by 
\begin{equation}
	L_{RH} \sim \sqrt{ \frac{U}{2 \Omega R \cos \phi} }
\end{equation} 
where $U$ is the characteristic wind speed, $R$ is the radius, $\Omega = 2\pi / P$ where $P$ is the period, and $\phi$ is the latitude of the atmospheric feature. Assuming that the wind speeds and latitudes are the same then $L_{RH} \sim \sqrt P$. Thus we would expect the maximum atmospheric feature size to increase with longer rotational periods, explaining the increasing variability amplitude with period in Figure \ref{fig:per_amp_j}. \added{ Beyond periods of $9~$hr, this correlation does not seem to hold. This suggests that that periods greater than $\sim7-9~$hr, the Rhines length is no longer the dominant factor in controlling the size of atmospheric features.}

\begin{figure}[tb]
	\centering
		\includegraphics[scale=0.5]{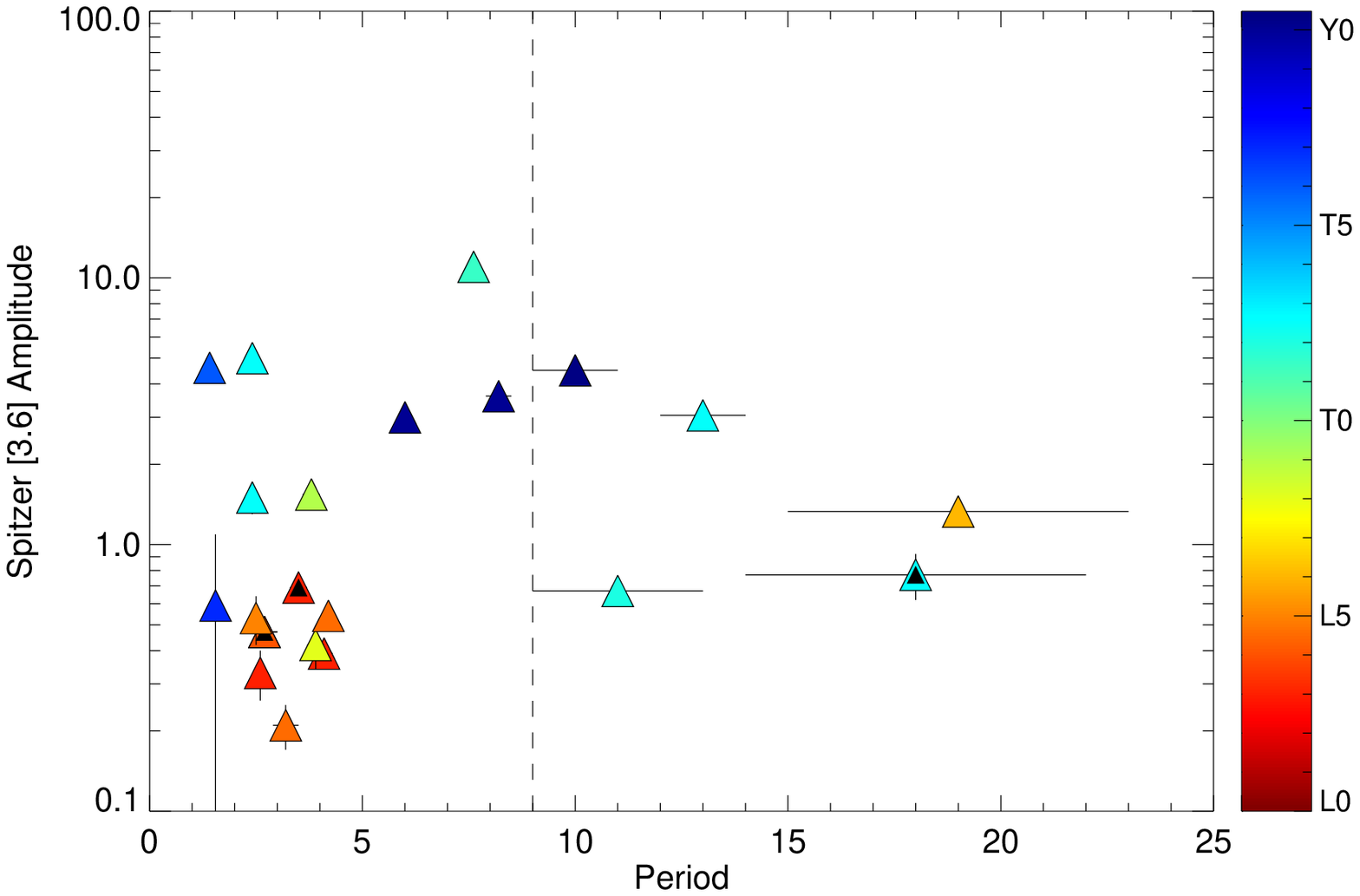}
	\caption{Same as Figure \ref{fig:per_amp_j}, but showing \textit{Spitzer} variability detections. \deleted{We calculate a $p$-value $\sim6\%$ for the correlation between variability amplitude and rotational period for periods $<9~$hr.}}
	\label{fig:per_amp_s}
\end{figure}

Figure \ref{fig:per_amp_s} shows the \textit{Spitzer} amplitudes plotted against rotation periods for all \textit{Spitzer} variable objects with published periods (presented in Table \ref{table:Spitzer}). \textit{Spitzer} observations are in general longer than ground-based $J$-band observations (\citet{metchev2015a} employed $\sim20~$hr observations for their \textit{Spitzer} survey) and are thus sensitive to longer periods. \textit{Spitzer} lightcurves have much higher photometric precision than ground-based studies and thus are also sensitive to lower amplitudes. However, clearly mid-IR variability is intrinsically lower than the near-IR variability. \replaced{For periods $<7-9~$hr, the variability amplitude seems to increase with increasing periods, as we saw for the $J$-band detections. This relation is significant at the $93\%$ level, so is more tentative than the $J$-band result.}{In contrast to the $J$-band data, Kendall's $\tau$ produces $p$-value $\sim80\%$, thus we find no correlation between variability amplitude and rotation period in this case. }
At longer periods, the observed variability amplitudes appear to decrease, \replaced{, something that is not observed in the $J$-band.}{however the sparse number of data points prevents us from confirming this.} \deleted{This suggests that that periods greater than $\sim7-9~$hr, the Rhines length is no longer the dominant factor in controlling the size of atmospheric features.} The highest variability amplitudes in the mid-IR case are detected in the late T's and early Y's, in contrast to the $J$-band, where high amplitudes are detected in L/T transition objects. Again, the young L-type objects may have slightly enhanced amplitudes when compared to field L-type brown dwarfs \citep{metchev2015a}.

\subsection{Investigating Colour Anomalies of the Sample}

\begin{figure}[tb]
	\centering
		\includegraphics[scale=0.51]{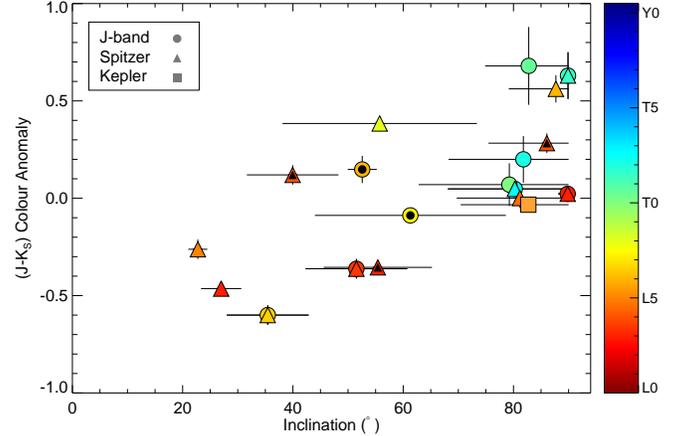}
	\caption{ Colour anomaly plotted against inclination for sample in Table \ref{tab:sample}. Young objects are denoted by a black inset. }
	\label{fig:inc_colan}
\end{figure}

We define the colour anomaly of each object as the median 2MASS $J-K_S$ colour subtracted from the $J-K_S$ colour of the object. Median colours for L0 - T6 objects were taken from \citet{Schmidt2010}. For 2M0050, the T7 object, we calculated the median of all IR T7 objects from DwarfArchives.org (20 objects) and found the median T7 $J-K_S$ colour to be $-0.04\pm 0.43$. This is a much higher error than those in \citet{Schmidt2010} and was thus left out of the analysis. With no $J-K_S$ measurement of Y dwarfs, it was not possible to include WISE0855, WISE1405 and WISE1738. \citet{Liu2016} provides linear relations between spectral type and absolute magnitude for {VL-G}  and {INT-G}  brown dwarfs, and these were used to calculate the median colours for the low-gravity sample.

Figure \ref{fig:inc_colan} displays the colour anomaly of objects listed in Table \ref{tab:sample} plotted against their inclinations. We note a correlation between the $J-K_S$ colour anomaly and inclination whereby objects viewed equator-on appear redder than objects viewed at lower inclinations.

Calculating the correlation coefficient and $p$-value, we find that the relation between colour anomaly and inclination is statistically significant with a $p$-value \replaced{$<1\%$}{$=0.4\%$}.  Objects we observe to be redder than the median are equator-on, whereas objects appearing bluer than the median are closer to pole-on. This result could be interpreted by the idea first proposed by \citet{Kirkpatrick2010}, that viewing angle determines the spectral appearance of a brown dwarf. This could occur if clouds are not homogenously distributed in latitude or if grain size and cloud thickness vary in latitude. Our results can be explained if thicker or large-grained clouds are situated at the equator, while thinner or small-grained clouds are situated at the poles.

\begin{figure}[tb]
	\centering
		\includegraphics[scale=0.51]{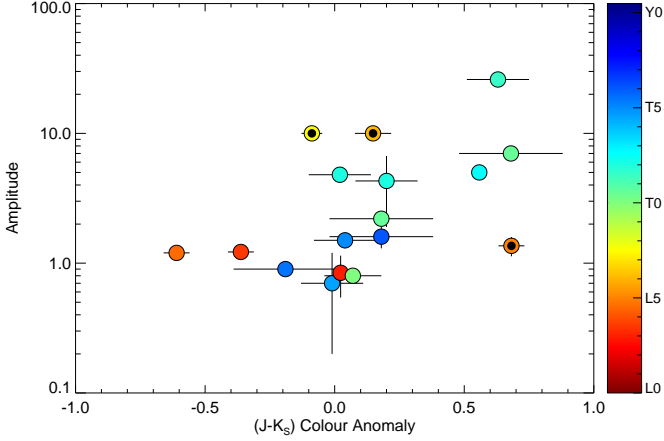}
	\caption{Amplitude plotted against colour anomaly for $J$-band variability detections.  }
	\label{fig:amp_colan_j}
\end{figure}

\begin{figure}[tb]
	\centering
		\includegraphics[scale=0.51]{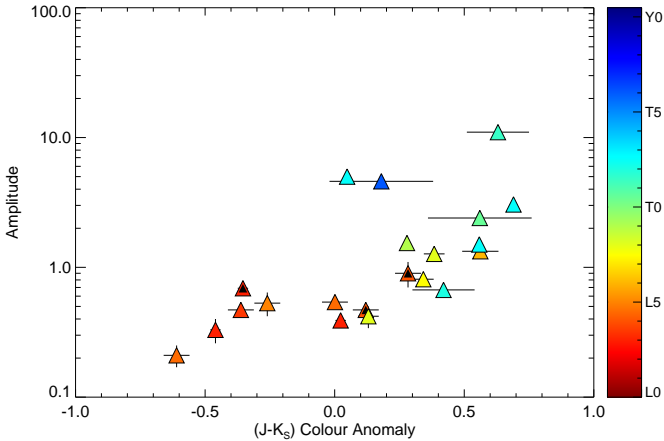}
	\caption{Amplitude plotted against colour anomaly for \textit{Spitzer} variability detections.  }
	\label{fig:amp_colan_s}
\end{figure}

Figures \ref{fig:amp_colan_j} and \ref{fig:amp_colan_s} show the variability amplitude plotted against the colour anomaly for $J$-band and \textit{Spitzer} detections respectively. Both plots exhibit a consistent trend, whereby field objects that are redder than the median display higher $J$-band and \textit{Spitzer} variability amplitudes. The field objects with the highest observed variability amplitudes are those with the reddest $J-K_S$ colours of their spectral type. We find that this correlation is significant at the \replaced{$95\%$}{$93\%$} and $99\%$ levels for the $J$-band and mid-IR detections respectively.
This relation may be explained by consideration of viewing angle. If redder brown dwarfs are equator-on, and equator-on objects exhibit the highest amplitudes, then it follows that redder brown dwarfs should display the highest variability amplitudes. Similarly, bluer brown dwarfs are viewed close to pole-on, so the observed variability amplitude will be reduced due to the viewing angle.

\added{We also see trends related to spectral type in both figures which could explain the observed relation. In the $J-$band case (Figure \ref{fig:amp_colan_j}), the early to mid-L spectral type field dwarfs display a blue anomaly while the L/T transition field dwarfs display a red $(J-K_S)$ colour anomaly. The late T type objects with detected variability display colours that are relatively close to the median.
These trends  are shown even more clearly for the \textit{Spitzer} detections (Figure \ref{fig:amp_colan_s}). The low-amplitude variability detections are observed in early L type brown dwarfs displaying a blue anomaly. We observe higher amplitude variability in L/T transition objects that display a red anomaly. This trend could be explained by variability due to the breakup of silicate clouds. L type brown dwarfs with thick silicate clouds generally appear red, while the  relatively cloudless T dwarfs  appear more blue. Thus, L dwarfs whose clouds have begun to break up will appear bluer than the median, and produce variability due to these patchy clouds. On the other hand, early T dwarfs that still have clouds in their atmospheres will appear redder than the median, resulting in photometric variability as these clouds rotate in and out of view. While this simple idea is an attractive explanation, spectroscopic  variability observations have shown that cloud evolution in L and T brown dwarfs atmospheres is significantly more complicated than simple formation of cloud holes \citep{Apai2013,Buenzli2012,Buenzli2015a,Yang2016}.

Furthermore, we see that surface gravity has an effect on this relation in both bands. For the $J-$band detections (Figure \ref{fig:amp_colan_j}), the low-surface gravity objects do not seem to follow the trend in spectral type, and appear among the L/T transition  field objects. It seems that low-surface gravity objects that are redder than the median appear variable but with only three detections we cannot confirm this. In contrast, for the \textit{Spitzer} detections, $2/3$ of the low-surface gravity objects seem to follow the overall trend, with one object falling closer to the L/T transition field brown dwarfs. Variability surveys of young, low-surface gravity objects will clarify these possible deviations from the field brown dwarf population. }


\section{Summary and Conclusions}

In this paper we explored the effects of inclination angle on measured variability amplitudes and whether brown dwarfs display similar intrinsic amplitudes. We further went on to examine the relation between inclination angle and spectral appearance. We determined the inclination angle of 19 variable brown dwarfs using archival Keck data and estimates on radius. We analyse the full sample of L, T and Y spectral type brown dwarfs with published $J$-band and \textit{Spitzer} variability detections. 

We conclude that brown dwarfs have different intrinsic amplitudes, dependent on properties such as spectral type, rotation period and surface gravity. In this paper we find evidence that the variability amplitude may increase with rotational period for periods $<7-9~$hr. \replaced{This result is statistically significant for $J$-band variabilty detections and more tentative for mid-IR detections.}{This result is significant at the $93\%$ level for $J-$band detections but is not significant for \textit{Spitzer} detections.}
The inclination angle affects the observed amplitude due to a projection effect as well as  atmospheric attenuation. Our toy model suggests that $J$-band variability is more strongly affected by inclination when compared to \textit{Spitzer} variability.  This may be due to the $J$-band probing deeper levels in the atmosphere. This results in the flux coming from these deeper levels being attenuated more due to increased path lengths at lower inclinations. All brown dwarfs with mid-IR variability detections are inclined at an angle $>20^{\circ}$. In the near-IR, we find that all brown dwarfs with $J$-band variability detections are inclined at an angle $>35^{\circ}$.

 We find a trend between the colour anomaly and inclination of our sample that is statistically significant at the $99\%$ level. Field objects viewed equator-on appear redder than the median for their spectral type, whereas objects viewed at lower inclinations appear bluer. This supports the idea that our viewing angle influences the spectral and photometric appearance of a brown dwarf. These results can be explained if thicker or large-grained clouds are situated at the equator, with thinner or small-grained clouds at the poles.
 We also find a strong correlation between colour anomaly and both mid-IR and $J$-band variability, where redder objects have higher variability amplitudes. This again suggests that the spectral appearance of a brown dwarf is strongly affected by its inclination angle.


\acknowledgments

The authors would like to thank Jack Gallimore for his contributions to the fitting code. JMV acknowledges the support of the University of Edinburgh via the Principal's Career Development Scholarship. KNA acknowledges support from the Isaac J. Tressler Fund for Astronomy at Bucknell University. BB gratefully acknowledges support from STFC grant ST/M001229/1.

\bibliography{Full}


\end{document}